# NUCLEAR CLUSTERING AND INTERACTIONS BETWEEN NUCLEONS

OLIVER K. MANUEL

*Chemistry Department, University of Missouri*
*Rolla, Missouri 65401, U.S.A.*

Nuclear mass data is used to show evidence of clustering of nucleons, attractive n-p interactions, and repulsive but symmetric n-n and p-p interactions after correcting for the repulsive Coulomb interactions of positive nuclear charges. These interactions suggest a possible source of energy in neutron stars, in second generation stars which formed on them, and demonstrate the need to develop a theoretical basis for understanding the fundamental problem of a) interactions between nucleons, and b) neutron-emission by penetration of the gravitational barrier surrounding a neutron star.

## 1 Introduction

My interest in obtaining a better understanding of the interactions between nucleons was motivated by an early appreciation for the beautiful set of data represented by masses of the nuclides while still a graduate student in the 1960's, and a gradual realization over the ensuing decades that some process other than hydrogen-fusion might be producing some of the Sun's energy. Finally on Christmas day of 2000, three students and I submitted a report to the Foundation for Chemical Research, Inc. [1] with a summary of information obtained when we abandoned the conventional approach and used something akin to the reduced variables in van der Waals' equation of corresponding states to study properties of the 2,850 nuclides tabulated in the latest report from the National Nuclear Data Center [2].

Trends in the reduced variables, Z/A or charge per nucleon, and M/A or potential energy per nucleon, revealed evidence that the n-p interactions are attractive, while the n-n and p-p interactions are repulsive and symmetric after correcting for the well-known repulsive Coulomb interactions between positive nuclear charges. These findings were unexpected from the two-nucleon interaction potentials described in standard nuclear textbooks. Nuclear disintegration, far from the valley of beta stability, were attributed to proton and neutron drip lines beyond which "the unbound proton or neutron drips out of the nucleus." [3, page 381].

On the contrary, trends in the empirical data indicate that neutron or proton emission releases large amounts of energy if the parent nuclide is far from the valley of beta stability. Proton emission releases the largest amount of energy when Z/A ~ 1.0, but this probably does not correspond to any natural form of matter heavier than $^1H$. However, neutron emission when the parent nuclide has Z/A ~ 0, e.g., a neutron star, typically releases 10-22 MeV per neutron emitted. This converts a larger fraction of rest mass into energy than fission or fusion. Thus, neutron emission may account for a large fraction of the energy released by the Sun and other stars that formed out of fresh supernova debris [4, 5].

In an effort to encourage others to address and to develop a theoretical basis for understanding this fundamental problem, empirical evidence [2] for nuclear clustering in the interactions between nucleons was presented at the 6th Workshop on Quantum Field Theory Under the Influence of External Conditions at the University of Oklahoma. This will be summarized below to show the need for a better understanding of a) nucleon interaction potentials, and b) neutron penetration of the gravitational potential barrier of a neutron star. The results may advance nuclear physics and our understanding of the source of energy that bathes planet Earth and sustains life.

**2 Nuclear Clustering and Interactions between Nucleons**

The "cradle of the nuclides", Figure 1, illustrates major trends when data for ground states of the 2,850 known nuclides [2] are plotted in terms of Z/A, charge per nucleon, versus M/A, mass or total potential energy per nucleon, and then sorted by mass number, A. All nuclides have values of 0=Z/A=1, and these define a cradle shaped like a trough made by holding two cupped hands together. The more stable nuclides lie along the valley, and $^{56}Fe$ lies at the lowest point. Lighter, more fusible nuclides occupy higher positions, up the steep slope to the left of A = 56 in Figure 1. The heavier, more fissionable nuclides occupy slightly higher positions, up the gradual slope to the right of A = 56.

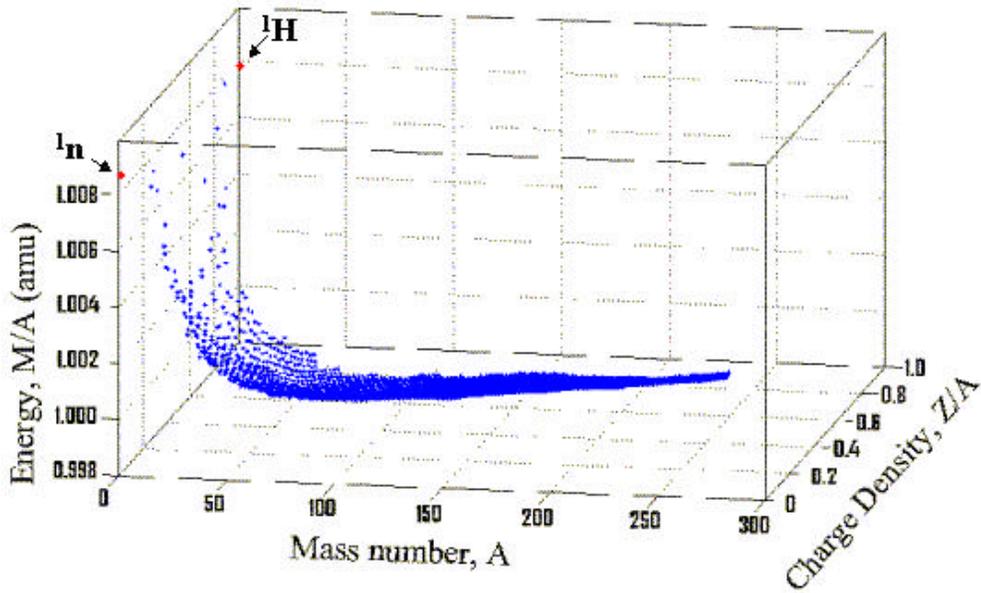

Fig. 1. The cradle of the nuclides.

At any given value of A, the the masses of the nuclides define a "mass parabola" as the values of Z/A increase from the lowest known value, closest to Z/A = 0, to the highest known value closest to Z/A = 1. The most stable charge on any nuclide of mass number A generally lies about midway between the front and back planes in Figure 1, at the low point in the mass parabola.

Nuclides that are closer to the front plane, i.e., those having lower values of Z/A, tend to decay by negatron (electron) emission; nuclides that are closer to the back plane in Figure 1, i.e., those having higher values of Z/A, tend to decay by positron emission or electron capture. There is a minor "saw-tooth" fine-structure caused by even-even versus odd-odd effects when A is an even number. To avoid this distraction, the next three graphs will show these trends in more detail when A is an odd number.

Figure 2 shows for example a cross section through Figure 1 at A = 27. The low point in the mass parabola occurs at $^{27}$Al. From left to right, all eight known nuclides [2] at A = 27 are $^{27}$F, $^{27}$Ne, $^{27}$Na, $^{27}$Mg, $^{27}$Al, $^{27}$Si, $^{27}$P, and $^{27}$S.

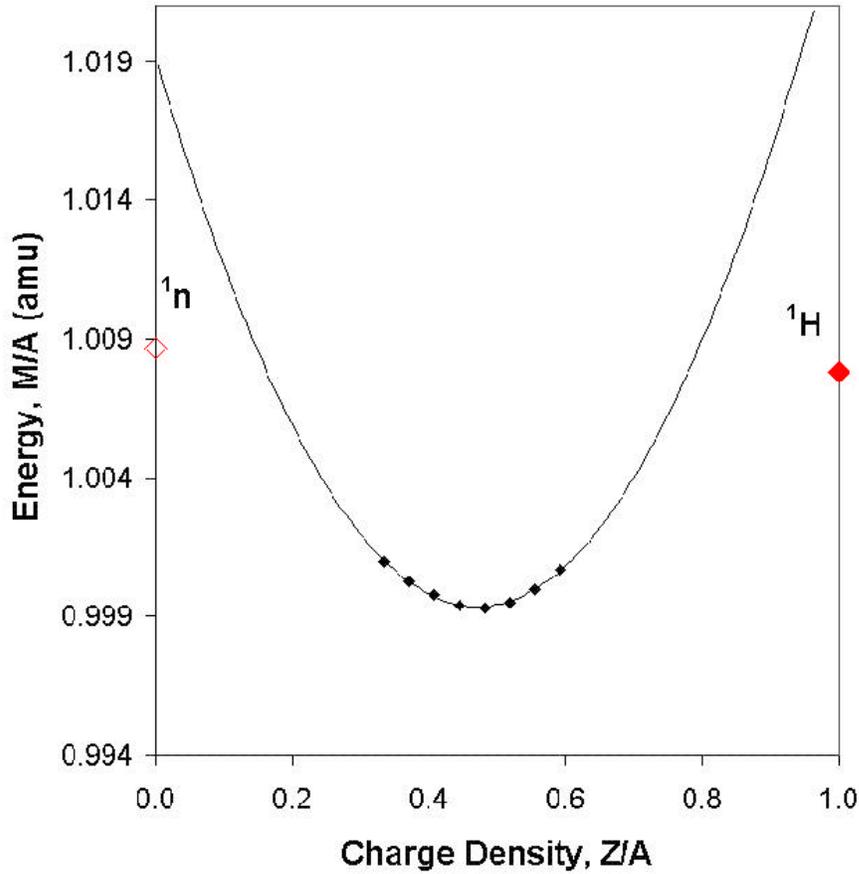

Fig. 2. A cross section through Fig. 1 at A = 27.

Figure 2 also shows values of M/A for unbound nucleons, a neutron on the left and a $^1$H atom on the right at Z/A = 0 and Z/A = 1.0, respectively. The empirical mass parabola defined by $^{27}$F, $^{27}$Ne, $^{27}$Na, $^{27}$Mg, $^{27}$Al, $^{27}$Si, $^{27}$P, and $^{27}$S yields much higher values of M/A for an assemblage of 27 neutrons at Z/A = 0 and for an assemblage of 27 protons or Z/A = 1.0, respectively. Cross-sectional cuts through Figure 1, at any value of A >1, reveal an empirical mass parabola with values of M/A > M($^1$n) at Z/A = 0 and values of M/A > M($^1$H) at Z/A = 1.0.

Typically the excess energy associated with these assemblages of pure neutrons or protons is 10 MeV per nucleon, plus energy from Coulomb repulsion at Z/A = 1. Unlike the imagined release of neutrons from a nucleus near the neutron drip line [3, page 381], repulsion between neutrons may cause neutron emission and the release of energy from a neutron star.

Coulomb repulsion between positive nuclear charges contributes to the high value of M/A for an assemblage of 27 protons on the right side of Figure 2, but not to a nucleus of 27 neutrons on the left.

In fact the difference between values of M/A, obtained at the intercepts where $Z/A = 1.0$ and $Z/A = 0$, arises from Coulomb repulsion and increases linearly with $A^{2/3}$ over the mass range, $A = 1-41$ [6]. The slope of this line is indistinguishable from that defined by the familiar β- decay of mirror nuclei close to the line of β- stability, e.g., ($^{1}$H, $^{1}$n), ($^{3}$He, $^{3}$H), ($^{5}$Li, $^{5}$He), ($^{7}$Be, $^{7}$Li,), . . . , ($^{41}$Sc, $^{41}$Ca) [7, page 35].

Thus, the values obtained for M/A from empirical mass parabolas at $Z/A = 1.0$ and $Z/A = 0$ yield the same nuclear radius and the same coefficient for the Coulomb energy term as the mirror nuclei close to the line of β- stability for $A = 1-41$ [6].

The decay energy, and hence the Coulomb energy of heavier nuclides, A>41, can also be obtained from differences indicated by mass parabolas for values of M/A at $Z/A = 1.0$ and $Z/A = 0$. Figure 3 shows the results for all odd values of A, from $A = 1$ to $A = 263$.

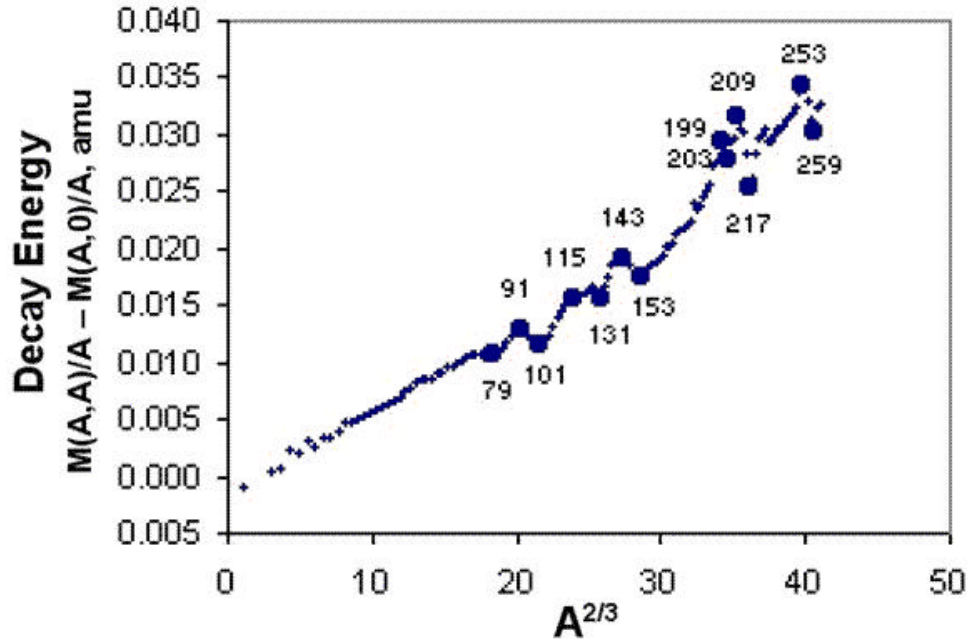

Fig. 3. Decay energies of extreme nuclides, where Coulomb energy drives $Z/A = 1 \longrightarrow Z/A = 0$, for all odd values of A from $A = 1$ to 263.

The decay energies of light nuclides in Figure 3 vary linearly with $A^{2/3}$, but fine structure starts to appear near $A \sim 80$. Peak energies, at $A = 91, 115, 143, 199, 209$ and $253$, likely arise from high Coulomb energy at $Z/A = 1$ because of clustering of nucleons into tightly packed structures. Likewise, valleys at $A = 79, 101, 131, 153, 203, 217$ and $259$ likely mean low Coulomb energy at $Z/A = 1$ because of more loosely packed nucleons.

There is no Coulomb energy associated with the other extreme form of nuclides, at Z/A = 0. These are the intercepts of mass parabolas with the front plane in Figure 1 at each value of A. However, they also reveal fine structure, as shown in Figure 4 for all odd values of A from A = 1 to 263.

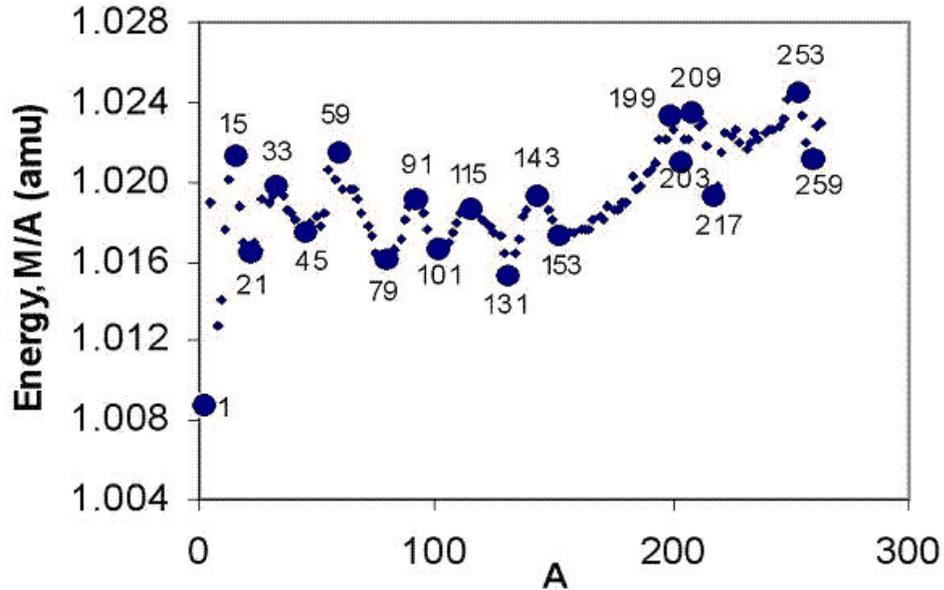

Fig. 4. Values of M/A at Z/A = 0 for all odd A parabolas, A = 1 to 263.

The data in Figure 4 includes, for example, M/A = 1.019 at A = 27, as shown earlier in Figure 2. Note that all values of M/A are higher than that of the free neutron (A = 1) for A>1. This was recognized as an indication of repulsive interactions between neutrons in 2000 [1]. Neutron emission from these nuclides would typically generate about 10 MeV per nucleon, as shown by the example in Figure 2 for A = 27.

The rhythmic distribution in values of M/A with A at Z/A = 0 was not understood in 2000. However, the peaks and valleys in Figure 4 occur at the same mass numbers as those in Figure 3 for A≥79. Nuclear clustering into tightly packed structures produces peaks at A = 91, 115, 143, 199, 209 and 253 in Figure 3 from enhanced Coulomb repulsion. Nuclear clustering into tightly packed structures produce peaks at these same mass numbers in Figure 4 from enhanced repulsion between neutrons. Loosely packed nucleons produce valleys at A = 79, 101, 131, 153, 203, 217 and 259 in Figure 3 from reduced Coulomb repulsion between loosely packed protons and in Figure 4 from reduced repulsion between loosely packed neutrons.

The rhythmic scatter of data in Figure 4 suggests that nuclear clustering also occurs below A = 79. However, the positive charge apparently maintains a spherical shape. Thus, the Coulomb energy is proportional to $A^{2/3}$ at A<79 in Figure 3, as well as in ordinary mirror nuclides [7, page 35].

**3 Importance of a Theoretical Basis for Nucleon Interactions**

Indications of the need for a better theoretical understanding of interactions between nucleons and neutron-emission first surfaced in the early 1960s when Fowler et al. [8] noted that some event at the birth of the solar system likely produced short-lived nuclides, deuterium, and the isotopes of lithium, beryllium and boron [1]. Subsequent analyses of meteorites in the 1970's suggested that the entire solar system may have formed from heterogeneous debris of a single supernova, with the Sun forming on the collapsed supernova core [4, 5] in the manner illustrated in Figure 5.

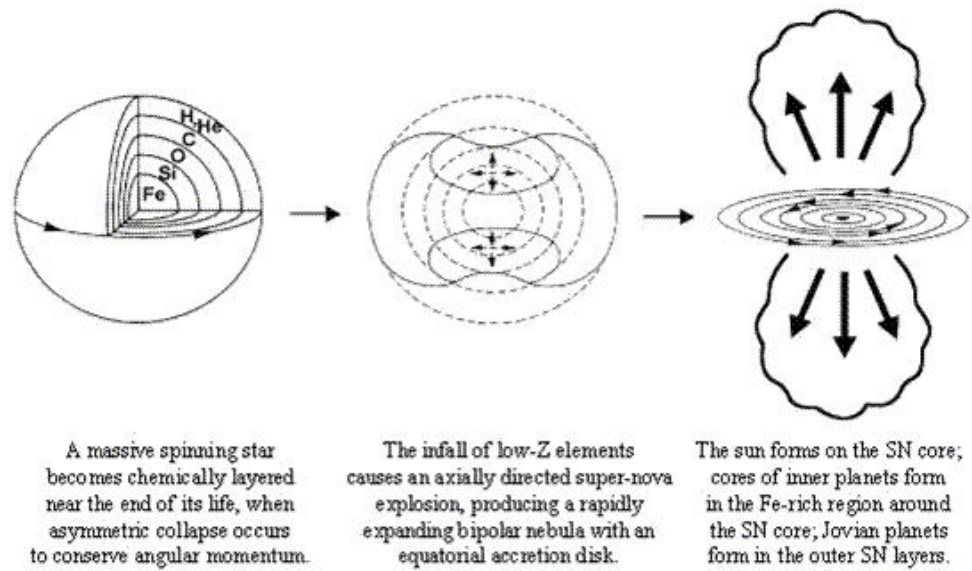

| A massive spinning star becomes chemically layered near the end of its life, when asymmetric collapse occurs to conserve angular momentum. | The infall of low-Z elements causes an axially directed super-nova explosion, producing a rapidly expanding bipolar nebula with an equatorial accretion disk. | The sun forms on the SN core; cores of inner planets form in the Fe-rich region around the SN core; Jovian planets form in the outer SN layers. |

Fig. 5. Formation of the solar system from a spinning supernova.

The scenario depicted in Figure 5 quickly gained support by findings that material in meteorites and planets [9, 10, 11, 12] have linked elemental and isotopic variations induced by stellar nuclear reactions. The hydrogen filled object at the center of the solar system was the main obstacle to this scenario. That was solved by the discovery that the interior of the Sun is made of the same elements that comprise 99% of the material in meteorites [13] and that mass separation in the Sun enriches lighter elements and the lighter isotopes of each element at the solar surface by a common mass fractionation power law [14]. Recent reviews [15, 16, 17] summarize these and other measurements since 1960 that support the scenario in Figure 5.

## 4 Conclusions

A better theoretical understanding of interactions between nucleons and penetration of the gravitational barrier around a neutron star is now needed to see if solar luminosity (SL) might reasonably be explained by the following sequence of events [17]:

- Neutron emission from a central neutron star ( >57% SL)

$$<{}_0^1n> \;\longrightarrow\; {}_0^1n \;+\; \sim 10\text{-}22 \text{ MeV}$$

- Neutron decay ( <5% SL)

$${}_0^1n \;\longrightarrow\; {}_1^1H^+ + e^- + \text{anti-}\nu \;+\; 0.782 \text{ MeV}$$

- Fusion and upward migration of H+ ( <38% SL)

$$4\, {}_1^1H^+ + 2\, e^- \;\longrightarrow\; {}_2^4He^{++} + 2\,\nu \;+\; 27 \text{ MeV}$$

- Escape of excess H+ in the solar wind (100% SW)

$$3 \times 10^{43} \text{ H}^+/\text{year} \;\longrightarrow\; \text{depart in the solar wind}$$

I will be happy to cooperate with anyone wishing to address and to develop a theoretical basis for understanding this fundamental problem.

## Acknowledgements

This study was supported by the University of Missouri-Rolla and the Foundation for Chemical Research, Inc., which kindly consented to our request to reproduce these figures from earlier reports to the Foundation for Chemical Research, Inc.